**A DECADE OF AGILE**

# Detection of terrestrial gamma-ray flashes with the AGILE/MCAL

Alessandro Ursi[1] · Martino Marisaldi[2] · Marco Tavani[1] on behalf of the AGILE Team

**Abstract**
AGILE is one of the satellites currently detecting terrestrial gamma-ray flashes (TGFs). In particular, the AGILE Mini-CALorimeter detected more than 2000 events in 8 years activity, by exploiting a unique sub-millisecond timescale trigger logic and high-energy range. A change in the onboard configuration enhanced the trigger capabilities for the detection of these events, overcoming dead time issues and enlarging the detection rate of these events up to > 50 TGFs/month, allowing to reveal shorter duration flashes. The quasi-equatorial low-inclination (2.5°) orbit of AGILE allows for the detection of repeated TGFs coming from the same storms, at the same orbital passage and throughout successive orbital overpasses, over the same geographic region. All TGFs detected by AGILE are fulfilling a database that can be used for offline analysis and forthcoming studies. The limited number of missions currently detecting these brief terrestrial flashes makes the understanding of this phenomenon very challenging and, in this perspective, the AGILE satellite played and still plays a major role, helping shedding light to many aspects of TGF science.

## 1 Introduction

Terrestrial gamma-ray flashes (TGFs) are sub-millisecond gamma-ray emissions, produced in Earth atmosphere, and correlated with thunderstorms and electric activity. Usually, TGFs last hundreds of µs, exhibiting energies from hundreds keV up to tens of MeV and representing the most energetic phenomenon naturally occurring on Earth. Up to now, these elusive events have been detected by high-energy astrophysics satellites, such as the compton gamma-ray observatory (CGRO) (Fishman 1994), the reuven ramaty high-energy spectroscopic imager (RHESSI) (Grefenstette et al. 2009;

Smith et al. 2010), the Astrorivelatore Gamma ad Imaggini LEggero (AGILE) satellite (Marisaldi et al. 2010, 2015), the Fermi space telescope (Briggs et al. 2010; Roj et al. 2019), and the BeppoSAX satellite (Ursi et al. 2017). Nevertheless, TGFs have been revealed from aircraft (Smith et al. 2011) and from ground facilities (Dwyer et al. 2012; Bowers et al. 2017), as well. The recently launched Atmosphere–Space Interaction Monitor (ASIM) mission, housed onboard the International Space Station (ISS), will contribute to shed light on these evens, by performing multi-wavelength observations of the phenomenon.

TGFs occur at are deeply linked with lightning activity: as a consequence, their geographic distribution generally cluster within the Tropics and over land regions, where lightning occur more frequently, following the InterTropical Convergence Zone (ITCZ). This correlation is strongly confirmed by a large fraction of TGFs found in close time association (< 500 µs) with radio-atmospheric (sferic) waves emitted by lightning discharges (Connaughton et al. 2010; Marisaldi et al. 2015).

The production mechanism of TGFs is still poorly understood: currently, these events are thought to be produced by the Bremsstrahlung of relativistic electron avalanches, accelerated in thunderstorm electric fields and then braked by interacting with atoms and nuclei in the atmosphere (Gurevich et al. 1992; Dwyer et al. 2013; Dwyer 2008).

## 2 AGILE MCAL as TGF detector

The AGILE MiniCALorimeter (MCAL) is an all-sky monitor, sensitive in the range 0.4–100 MeV, composed of independent CsI(Tl) scintillation bars, and organized into two orthogonal planes (Labanti et al. 2009). It works both as a trigger for the Silicon Tracker (ST) detector in the so-called GRID mode, providing data acquisitions whenever the ST is triggered, and as an independent instrument in the so-called BURST mode, self-triggering transients and acquiring photon-by-photon data with 2 µs high time resolution. This last configuration is managed by a fully configurable onboard trigger logic and identifies transients by analyzing data in several different Ratemeters (RMs), with different search integration time (SIT) windows and energy ranges (Argan et al. 2008).

MCAL turned out to be an extremely suitable detector to reveal TGFs, especially due to the onboard sub-millisecond (0.293 ms) trigger logic timescale, very sensitive to short-duration gamma-ray transients. As a large number of triggers can be ascribed to electronic noise produced inside MCAL itself, an algorithm is required to identify TGFs, by means of an offline selection strategy based on: the evaluation of the released energy, the hardness-ratio of the events, and the spatial distribution of photons in the detector. Generally, the offline TGF search algorithm selects events with energies whose counts are uniformly distributed throughout the detector scintillation planes: this technique allows to reject non-physical triggers and led to the first AGILE TGF catalog Marisaldi et al. (2014), consisting in 308 events, acquired in 28 months observation (available online at http://www.asdc.asi.it/mcaltgfcat/).

All TGF detectors are affected by dead time, consisting in the time during which a detector is busy in processing the previous detected event and not sensitive to successive input signals. As TGFs release a large number of counts, (fluence = 0.1 ph cm$^{-2}$) in very short time intervals (hundreds of µs), dead time represents an issue that should be carefully evaluated. In the so-called standard configuration, run from the launch of AGILE until March 2015, the MCAL onboard trigger capabilities were highly affected by the dead time induced by the anti-coincidence (AC) shields, acting as a veto for background charged particles, but preventing the detection of extremely short (< 100 µs) events as well. The dead time effect strongly depended on the fluence of the events: for instance, TGFs with $T_{50} < 100$ µs and fluences $F > 0.08$ cm$^2$ had dead time fraction larger than 50% (Marisaldi et al. 2014). This resulted in the detection of a TGF sample biased toward longer time durations, with respect to events detected by other satellites, such as RHESSI and Fermi.

For this reason, starting March 2015, MCAL was put in a new onboard enhanced configuration, which overcame the dead time issue, by inhibiting the onboard active AC shield veto for MCAL, but keeping the same selection criteria adopted for the previous search algorithm: this resulted in an enhancement of the TGF detection of about one order of magnitude (i.e., up to > 50 TGFs/month) and allowing for the detection of shorter duration TGFs, with $T_{50} < 100$ µs (Marisaldi et al. 2015). Moreover, this allowed also to find close time associations (< 500 µs) between TGFs and sferics detected by the World Wide Lightning Location Network (Hutchins et al. 2012), as the chance of finding precise matches between the two populations is inversely proportional to the TGF time duration (Connaughton et al. 2010).

The AGILE nearly equatorial orbit (± 2.5°) represents a unique feature among satellites currently detecting TGFs: this low-inclination orbit makes AGILE pass over thunderstorm regions at successive orbital passages, experiencing a negligible latitudinal shift (< 1°) at each orbital revolution, and allowing MCAL to monitor these active systems in time. This results in the detection of multiple TGFs produced by the same thunderstorms, revealing either during the same orbital passage or during successive overpasses over the same geographic region (Ursi et al. 2016): on the other hand, satellites with higher-inclination orbits (e.g., RHESSI with ± 38° and Fermi with ± 26°) can detect consecutive TGFs during the same orbital passage (Grefenstette et al. 2009; Stanbro et al. 2018), but experience larger latitudinal shifts at each orbital revolution, preventing the detection of multiple TGFs throughout successive overpasses.

MCAL TGFs have been cross-correlated with meteorological data acquired by geostationary satellites, to investigate the meteorological conditions that favored the production of TGFs and to characterize the stages of the evolution of storms associated with these events (Ursi et al. 2017).

## 3 The new TGF sample

Figure 1 shows the total TGF sample of 2209 events detected by the AGILE MCAL: 498 TGFs were detected in the standard configuration, in the period 2 March 2009–22 March 2015, whereas 1711 TGFs were detected in the enhanced configuration, in the period 23 March 2015–27 November 2017. The detection rate increased of about one order of magnitude, from ~ 6 TGFs/month to ~ 50 TGFs/month, whereas the geographic distribution follows the typical TGF distribution, clustering over continental regions. The energy distribution is also compatible with the previous detected one. On the contrary, the time duration distribution substantially departs from the former one, passing from a median time duration $T_{90} = 710$ μs to $T_{90} = 255$ μs, confirming the enhanced sensitivity of MCAL to short duration events.

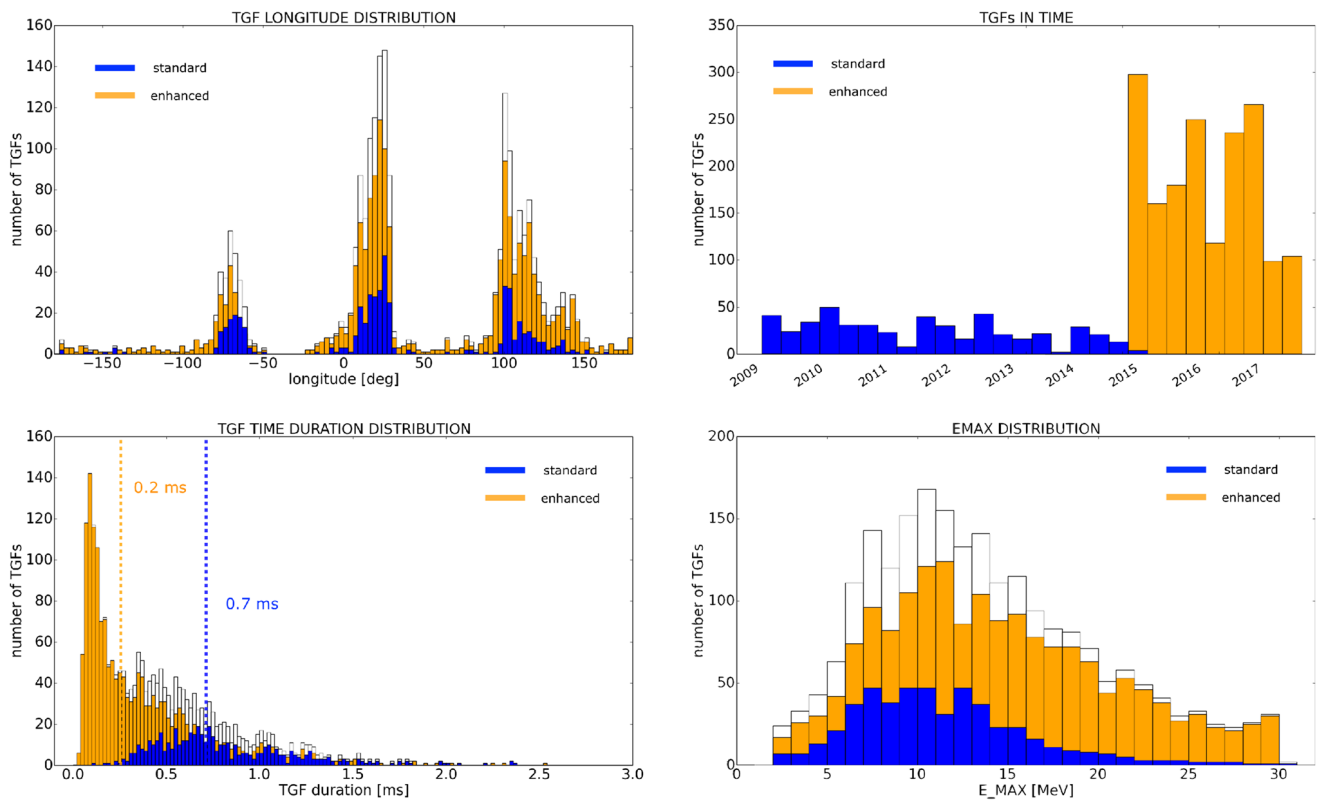

**Fig. 1** The total TGF sample (white) detected by the AGILE MCAL, as of 27 November 2017, with the related sub-samples detected in the standard (blue) and the enhanced (orange) configurations. The detection rate (**a**) increased to ∼ 50 TGFs/month, whereas the geographic (**b**) and energy (**c**) distributions are compatible with the previous detected one. On the contrary, the new configuration allowed for the detection of shorter duration events (i.e., median $T_{90} = 255$ μs) (color figure online)

Furthermore, MCAL is still detecting a large number of multiple TGFs, coming from within the same geographic region, as shown in Fig. 2. Among the 1711 TGFs collected in the enhanced configuration period, 106 events occurred within 1 min one from the successive one, representing multiple TGFs produced by the same storm and detected during the same orbital passage over a thunderstorm. Moreover, 58 multiple TGFs were detected at the first orbital resonance, 33 at the second orbital resonance, and 20 at the third orbital resonance, confirming the leading role of AGILE in the detection of these repeated events. Besides the main peaks due to orbital resonances, the waiting time distribution shows secondary features, made of both statistical fluctuations and the typical triple-peak geographic pattern of the most active regions in the world (i.e., Africa, Indonesia, and South America). Figure 3 shows an example of five multiple TGFs detected over Africa on 24-03-2015, either at successive overpasses over the same thunderstorm region, and during the same orbital passage, in a time interval of more than 8 h: the meteorological scenario is reconstructed by means of data acquired by the Meteosat-10 geostationary satellite, where colors refer to deep convective regions in the atmosphere.

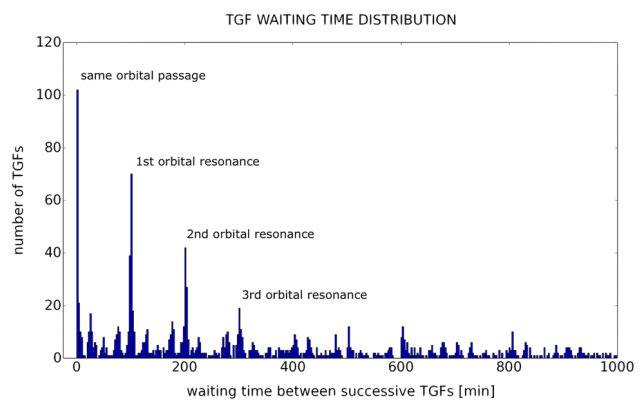

**Fig. 2** The waiting time distribution of MCAL TGFs, where peaks correspond to the *n*-th orbital resonances. These enhancements correspond to events coming from the same geographic region (bin = 1 min), either during the same overpass (first peak), or at the successive passages (successive peaks)

**Fig. 3** Example of five multiple TGFs (green stars) detected by the AGILE satellite overpassing the same thunderstorm region, in a time interval of more than 8 h. The first four events are detected at successive orbital passages, whereas the last one is detected after less than 1 min from the fourth one. The meteorological images are reconstructed using data of the Meteosat-10 geostationary satellite: colored regions correspond to deep convection in the atmosphere (color figure online)

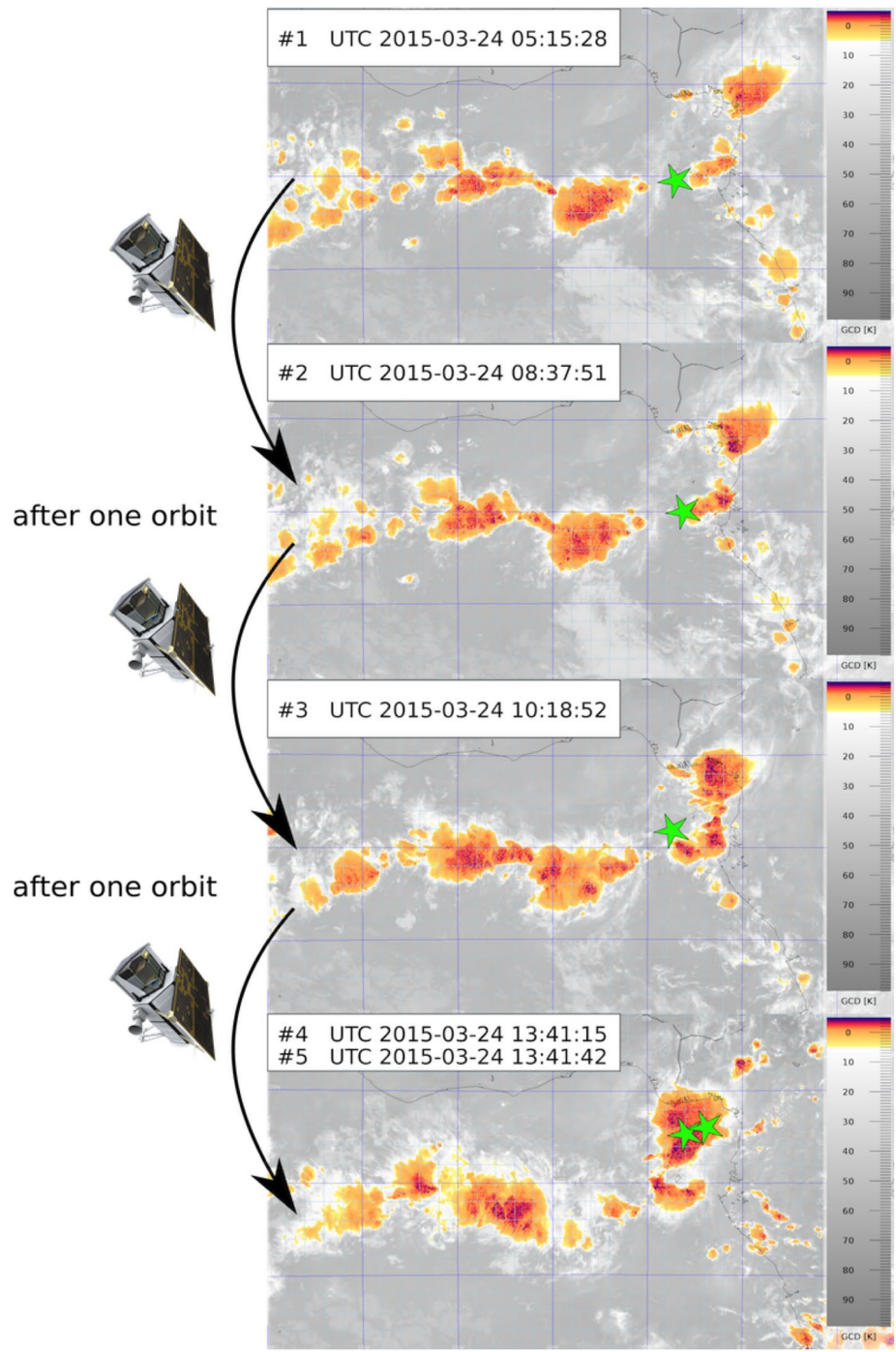

## 4 Conclusions

The AGILE MCAL detected a total of 2210 TGFs in 8 years activity. The largest fraction of these events (1711 TGFs) has been detected in the last 32 months, due to a new onboard trigger configuration, that enhanced the TGF detection rate up to more than 50 TGFs/month. The new TGF sample, collected in the period from 23 March 2015 to 27 November 2017, shows geographic and energetic distributions compatible with the sample acquired in the previous MCAL configuration, but a substantially different time duration distribution: the new configuration increased the detection capabilities of MCAL for shorter duration events, allowing to reveal events with durations down to tens of μs. Moreover, the new sample includes a large number of multiple TGFs, with tens of events detected either at the same orbital passage or at successive overpasses over the same active storm, confirming the unique leading role of the AGILE satellite in the detection of these repeated events.